\documentclass[12pt,a4paper]{article}

  \usepackage{color}
  \usepackage{a4wide}
  \usepackage{latexsym}
  \usepackage{epsf}
  \usepackage{amssymb,bbm}
  \usepackage{graphicx}
  \usepackage{amsmath, cite}
  \usepackage{amsmath,amssymb,amsthm}
  \usepackage{verbatim}
  \usepackage{hyperref}
  \usepackage[percent]{overpic}
  
    \usepackage{array}
\usepackage{etoolbox}
\patchcmd\thebibliography
 {\labelsep}
 {\labelsep\itemsep=0pt\relax} 
 {}
 {\typeout{Couldn't patch the command}}

  \newcolumntype{P}[1]{>{\centering\arraybackslash}p{#1}}
  
\newcommand{\f}[2]{\frac{#1}{#2}}

\newcommand{\dd}{\mathrm{d}}

\newcommand{\e}{\mathrm{e}}
\newcommand{\w}{\wedge}
\newcommand{\bbm}{\left(\begin{matrix}}
\newcommand{\ebm}{\end{matrix}\right)}
\newcommand{\bea}{\begin{eqnarray}}
\newcommand{\eea}{\end{eqnarray}}
\newcommand{\be}{\begin{equation}}
\newcommand{\ee}{\end{equation}}

\renewcommand{\cal}[1]{\mathcal{#1}}

\renewcommand{\d}{\textrm{d}}

\begin{document}
\numberwithin{equation}{section}

\begin{center}

\begin{flushright}
\end{flushright}
\vspace{1.5 cm}

{\LARGE {\bf The many faces of brane-flux annihilation}}  \\

\vspace{1.5 cm} {\large  Fridrik Freyr Gautason, Brecht Truijen and Thomas Van
Riet }\\
\vspace{0.5 cm}  \vspace{.15 cm} { Instituut voor Theoretische Fysica, K.U. Leuven,\\
Celestijnenlaan 200D B-3001 Leuven, Belgium}

\vspace{0.7cm} {\small \upshape\ttfamily  ffg, brecht.truijen, thomas.vanriet @fys.kuleuven.be
 }  \\

\vspace{1.2cm}

{\bf Abstract}
\end{center}

{\small Fluxes can decay via the nucleation of Brown-Teitelboim bubbles, but when the decaying fluxes induce D-brane charges this process must be accompanied with an annihilation of D-branes. This occurs via dynamics inside the bubble wall as was well described for $\overline{\text{D3}}$ branes branes annihilating against 3-form fluxes. In this paper we extend this to the other D$p$ branes with $p$ smaller than seven. Generically there are two decay channels: one for the RR flux and one for the NSNS flux. The RR channel is accompanied by brane annihilation that can be understood from the D$p$ branes polarising into D$(p+2)$ branes, whereas the NSNS channel corresponds to D$p$ branes polarising into NS5 branes or KK5 branes. We illustrate this with the decay of antibranes probing local toroidal throat geometries obtained from T-duality of the D6 solution in massive type  IIA. We show that $\overline{\text{D}p}$ branes are metastable against annihilation in these backgrounds, at least at the probe level.

}

\setcounter{tocdepth}{2}
\newpage

\section{Introduction}
The search for metastable SUSY-breaking flux vacua in string theory is of particular interest to phenomenology and remains an active area of research. Verifying that a solution is metastable requires the knowledge of the masses of all moduli and the decay mechanisms for topological data such as brane charge and flux. In this paper we focus on the latter and describe in general terms how fluxes and branes can annihilate against each other. Our work can be thought of as a generalisation of the decay of $\overline{\text{D3}}$ branes against 3-form fluxes of \cite{Kachru:2002gs} (see also \cite{Frey:2003dm, Brown:2009yb}) to the more general case where $\overline{\text{D}p}$ branes annihilate against either the NSNS 3-form flux $H_3$ or the RR $(6-p)$-form flux $F_{6-p}$. 
In this context \emph{antibranes} refer to the D-branes carrying a charge that is opposite to the background fluxes. In other words, the two terms on the right hand side of the Bianchi identity 
\be\label{Bianchi}
\d F_{8-p} = H_3\wedge F_{6-p} - 2\kappa^2_{10}Q_p\delta_{9-p}\,,
\ee 
have opposite orientation. The $\delta$-function form describes the brane sources. 

Following the original work of Brown and Teitelboim (BT) \cite{Brown:1987dd} we anticipate the presence of nonperturbative processes that lower the flux quanta. This process is the generalisation of Schwinger pair creation to spherical domain walls (BT bubbles) in spacetime that nucleate and can grow to infinite size if the vacuum inside the bubble has lower energy. 

For compact extra dimensions the integral of the LHS of equation (\ref{Bianchi}) must always vanish which implies that fluxes cannot decay unless the brane charges change in the same process. This suggests that the BT bubble is the thin wall limit of a more complicated process where the finite thickness of the bubble has to account for the dynamics that change the brane charge \cite{Danielsson:2014yga}. We can understand these dynamics by considering the infinitely thick limit. This involves a (generalised) Myers effect where the polarised brane can decay leaving less flux and changing the brane charge accordingly. As suggested these two effects are two distinct limits of a single process called brane-flux annihilation. This can be thought of as a closed string versus open string description of the same phenomenon.

An equivalent picture of brane-flux annihilation is that D$p$ branes can spontaneously materialise out of the flux. Those branes then annihilate with the antibranes and as a result there will be less antibranes and less flux. If the number of antibranes is smaller than the amount of nucleated D$p$ branes then this process leads to the creation D$p$ branes. 
It is the aim of this paper to explain how this phenomenon works for general branes and fluxes. 

This paper is organised as follows. 
We first set notations and context in section \ref{Notation} and then describe the brane-flux annihilation in the RR sector in section \ref{sec:RRdecay}. After reviewing the NS5 Wess-Zumino (WZ) action in type II supergravity, we describe the brane-flux annihilation in the NSNS sector. We then apply this to local throat geometries of the form $\mathbb{T}^{6-p}\times \mathbb{R}^3$ filled with fluxes that carry D$p$ charge. These geometries are obtained from T-dualising a solution found in \cite{Janssen:1999sa}. We demonstrate that, at probe a level, $\overline{\text{D}p}$ branes are metastable in these geometries. We end with conclusions in section \ref{Discussion}. In appendix \ref{Generalised} we comment on different types of brane-flux decay that are not discussed in the main text.

\section{Preliminaries and notation}\label{Notation}
Consider a compactification of type II supergravity which is supported by NSNS 3-form flux $H$ and RR flux $F_{6-p}$ on a compact space $M$ to some lower-dimensional maximally symmetric vacuum. Such solutions typically involve D$p$ branes and O$p$ planes to satisfy the RR tadpole condition
\be\label{tadpolecondition}
\int_{M} H_3\w F_{6-p} = 2\kappa^2_{10}Q_p~,
\ee
where $2\kappa^2_{10}=(2\pi)^7\alpha'^4$ and $Q_p$ denotes the total $p$-brane charge of the localized sources quantised according to
\be
Q_p = (2\pi)^{-p}(\alpha')^{-(p+1)/2}N_p\,,
\ee
where $N_p\in \mathbb{Z}$.

In order to have an explicit model in mind we assume that the compactification manifold is $9-p$ dimensional such that the sources are spacetime-filling and point-like in the compact dimensions. Our results are however valid for a general setup. Most widely know examples of this kind are IIB compactifications with 3-form fluxes to four dimensions \cite{Giddings:2001yu}, for which $p=3$. These can be generalised to $p=1,\ldots, 6$, see for instance \cite{Blaback:2010sj}. To simplify the presentation even further we pretend as if only two cycles are threaded with fluxes. There is the $A$-cycle of dimension $6-p$ and the $B$-cycle of dimension $3$ threaded by flux as follows:
\be
\int_{B} H_3 = (2\pi)^2 \alpha' K~, \qquad \int_{A}F_{6-p} = (2\pi)^{5-p} (\alpha')^{(5-p)/2} M~,
\ee 
with $K, M \in \mathbb{Z}$. Since the A and B cycle are each others dual we have from \eqref{tadpolecondition}
\be \label{tadpole}
N_p = KM\,.
\ee
In the rest of the paper we work in string units for which $2\pi\alpha'=1$.

Since we have both NSNS flux and RR flux there are two possible flux decay channels. Both of these channels proceed through the nucleation of a spacetime bubble where the inside of the bubble has less flux than the outside. This bubble will continue to grow if it is energetically favourable. The end state is either the true vacuum of the theory or another metastable state of lower energy. We recall the well-known Brown Teitelboim picture of flux decay and its generalisations before describing the stringy resolution of this process which is the topic of this paper.

The BT bubble that makes the NSNS flux decay is an NS5 brane wrapping the $A$-cycle and in the case of RR flux decay it is a D$(p+2)$ wrapping the $B$-cycle. One readily checks that these wrapped branes indeed have a single codimension in the ($p+1$)-dimensional vacuum. Only for $p=6$ is the NS5 is entirely inside noncompact spacetime\footnote{One could also envisage a situation in which the spherical brane is inside a compact manifold and then a flux cascade can take place as conjectured in \cite{Kleban:2011cs}. We leave an explicit string theory embedding of the cascade effect and its associated inflationary model \cite{DAmico:2012ji} for future investigations.}. The intersection diagram for these BT bubbles can be found in table \ref{table:Thin}.

\begin{table}[h!]
\begin{center}
\begin{tabular}{l m{2cm}|ccccc|ccccc|ccccc}
&Thin wall&  \multicolumn{5}{c|}{$p+1$} &  \multicolumn{5}{c|}{$A$-cycle}  &   \multicolumn{5}{c}{$B$-cycle} 
\\
\cline{2-17}
&O$p$/D$p$ && $\times$ &...& $\times$ &&&$-$&...&$-$&&&$-$&...&$-$ &
\\
NSNS decay:&NS5 && $\times$  &...&  $\uparrow$ &&&$\times$&...&$\times$&&&$-$&...&$-$&
\\
RR decay:&D$(p+2)$ && $\times$  &...& $\uparrow$&&&$-$&...&$-$&&&$\times$  &...& $\times$ &
\end{tabular}
\end{center}
\caption{\small The intersection diagram for the microscopic description of the BT bubbles that describe the thin wall limit of brane-flux annihilation. The arrow $\uparrow$ reflects the fact that the brane is moving in the radial spacetime direction. }
\label{table:Thin}
\end{table}

Because of the tadpole condition (\ref{tadpole}) a decay of the RR flux and NSNS flux must go together with a decay of brane charges as follows
\bea
\text{NSNS decay}\qquad:\qquad & K \rightarrow \,K-1 \quad &,\quad N_p \rightarrow N_p-M \,, \label{NSNSdecay}\\
\text{RR decay} \qquad:\qquad & M \rightarrow M-1 \quad &,\quad N_p \rightarrow N_p-K \,. \label{RRdecay}
\eea
Hence the BT bubble describes the thin wall limit of a more involved decay mechanism that should also account for the correct change in brane charge. Clearly, there should be some dynamics inside the bubble wall that accounts for the change in brane charge such that the tadpole condition is satisfied on both sides of the wall. To understand how the brane charge changes we discuss the thick wall limit. For the RR channel this is understood as the Myers effect \cite{Myers:1999ps} where some $\overline{\mathrm{D}p}$ branes polarise into a D$(p+2)$ brane filling spacetime and wrapping a trivial cycle inside the $B$-cycle. The NSNS channel is slightly more involved and requires a generalisation of the Myers effect where $\overline{\mathrm{D}p}$ branes polarise into an NS5 filling spacetime and wrapping a trivial cycle inside the $A$-cycle.\footnote{When $p=6$ this description does not make sense and we will speculate in section \ref{sec:D6} that this process involves a KK5 brane. } In both case the polarised brane can move either perturbatively or by the means of tunnelling inside the $A$- or $B$-cycle to a point where the wrapped trivial cycle pinches off. At this point the flux will have been lowered by one unit and the brane charge changes according to (\ref{NSNSdecay}, \ref{RRdecay}). The intersection diagram for this process can be found in table \ref{table:Thick}.

\begin{table}[h!]
\begin{center}
\begin{tabular}{l m{2cm}|ccccc|ccccc|ccccc}
&Thick wall&  \multicolumn{5}{c|}{$p+1$} &  \multicolumn{5}{c|}{$A$-cycle}  &   \multicolumn{5}{c}{$B$-cycle} 
\\
\cline{2-17}
&O$p$/D$p$ && $\times$ &...& $\times$ &&&$-$&...&$-$&&&$-$&...&$-$ &
\\
NSNS decay:&NS5 && $\times$  &...&  $\times$ &&&$\times$&...&$\uparrow$&&&$-$&...&$-$&
\\
RR decay:&D$(p+2)$ && $\times$  &...& $\times$&&&$-$&...&$-$&&&$\times$  &...& $\uparrow$ &
\end{tabular}
\end{center}
\caption{\small The intersection diagram for brane-flux decay in the thick wall limit. The arrow $\uparrow$ reflects the fact that the brane is moving in the corresponding cycle. }
\label{table:Thick}
\end{table}


In what remains we will discuss the dynamics in the thick wall approximation. We first analyse the decay of RR flux since it proceeds through the well known Myers effect \cite{Myers:1999ps}. In the subsequent section we analyse the more involved NSNS channel.

\section{The decay of RR flux}\label{sec:RRdecay}

\begin{figure}
\begin{center}
\begin{overpic}[width=0.3\textwidth,tics=10]{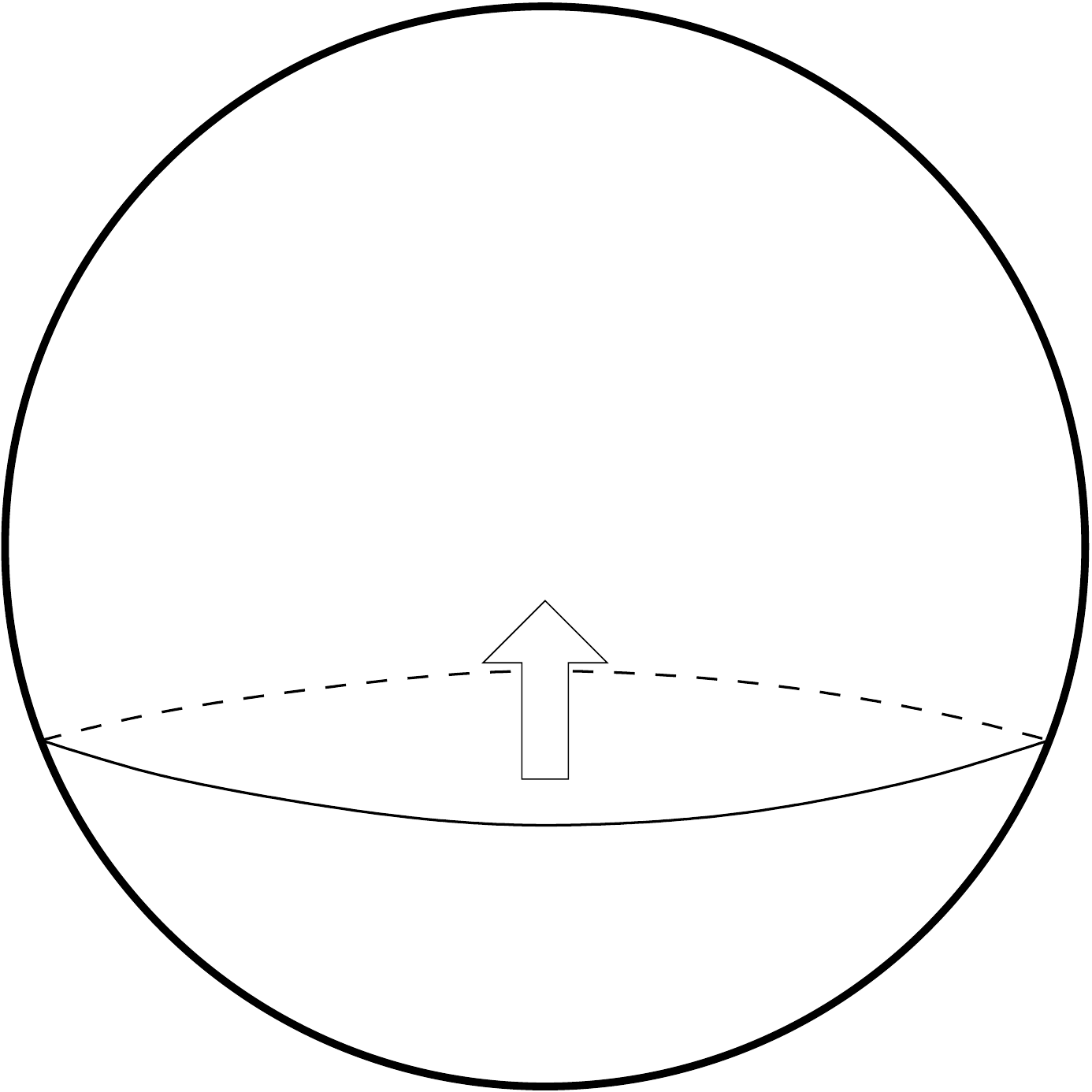}
\put(33,15){D$(p+2)$}
\put(-15,90){$B$-cycle}
\end{overpic}
\caption{\label{OurMyers}\small A D$(p+2)$ brane that wraps a trivial $S^2$ inside the spherical $B$-cycle. The D$(p+2)$ 
carries non-trivial worldvolume flux proportional to the volume form of the $S^2$ which ensures that the brane
carries $\overline{\mathrm{D}p}$ charge.}
\end{center}
\end{figure}

The BT bubble that decays a unit of RR flux corresponds to a  D$(p+2)$ brane that wraps the three-dimensional $B$-cycle as explained above. The tension of this domain wall is roughly given by
\be\label{thinsigma}
T = g_s^{-1}\mu_{p+2}\text{Vol}(B)\,,
\ee
in string units. This process must go together with the annihilation of some brane charges and we now explain that this 
occurs through the polarisation of 
D$p$ branes into D$(p+2)$ branes wrapping a trivial $S^2$ inside $B$ as depicted in figure \ref{OurMyers}. From here on 
we assume that $B$ has spherical topology. In case the topology is different the process will involve tachyon condensation as 
we explain in section \ref{NSdecay}. This D$(p+2)$ brane is not to be confused with the BT domain wall since the latter is a 
D$(p+2)$ that wraps the whole three dimensional $B$-cycle. 

Brane polarisation, a.k.a.~the Myers effect \cite{Myers:1999ps}, is enabled by the Wess-Zumino (WZ) coupling of the D$(p+2)$ brane
\be\label{DpWZ}
(-1)^p\mu_{p+2}\int \sigma(C_{p+3} + {\cal F}_2\w C_{p+1})~,
\ee
where the symbol $\sigma$ is the operator that reverses all indices and induces only minus signs in such a way that the Bianchi identity always has the same sign for the localised source, and
\be
{\cal F_2} = F_2 - B_2~,
\ee
with $F_2$ the worldvolume field strength living on the brane. The second term in the action can induce D$p$ charges. 

We define the position of the spherical D$(p+2)$ by the polar angle $\Psi$ on the 3-sphere. When $\Psi=0$ the spherical D$(p+2)$ pinches off and corresponds to some number $n$ of the original D$p$ branes. From the WZ action (\ref{DpWZ}) we can read off that the charge (measured in units of antibrane charge) induced by the spherical brane is 
\be
Q(\Psi) = \f{1}{2\pi}\int_{S^2(\Psi)} {\cal F}_2\,,
\ee
where the factor of $2\pi$ comes from comparing $\mu_{p+2}$ with $\mu_p$.
We are mainly interested in the difference between the charge at the South Pole $Q(0)$ and the North Pole $Q(\pi)$. Without knowing the actual profile of $B_2$ we can already deduce that
\be
\int_{S^2(\Psi\to\pi)}B_2-\int_{S^2(\Psi\to0)}B_2 =  \int_B H_3 = 2\pi K\,.
\ee
We can always consider a gauge of $B_2$ such that
\be
\int_{S^2(\Psi\to 0)}B_2 =0\,.
\ee
In order to induce $n$ antibrane charges at the South pole we  must have that 
\be
\int_{S^2(\Psi\to 0)}F_2 =2\pi n\,.
\ee
This is achieved when the worldvolume gauge field has a monopole configuration
\be
F_2 =\f{n}{2}\sin\theta\d\theta\wedge\d\varphi\,.
\ee
We assume this to remain unaltered throughout the motion along $\Psi$, see however \cite{Danielsson:2015eqa}. 
Hence we deduce that
\be
Q(\pi) = n-K\,.
\ee
So if the spherical D$(p+2)$ brane moves from the South Pole towards the North Pole it induces first $n$ $\overline{\text{D}p}$ charges and in the end $(K-n)$ D$p$ charges such that effectively $K$ D$p$ branes nucleated from the flux cloud and annihilated with $n$ $\overline{\text{D}p}$ branes. The spherical D$(p+2)$ does not carry any monopole charge but it does source $F_{6-p}$ locally which indeed causes the flux to decay ($M \rightarrow M-1$) as is required by the tadpole condition. 

\begin{figure}
\begin{center}
\includegraphics[height=50mm]{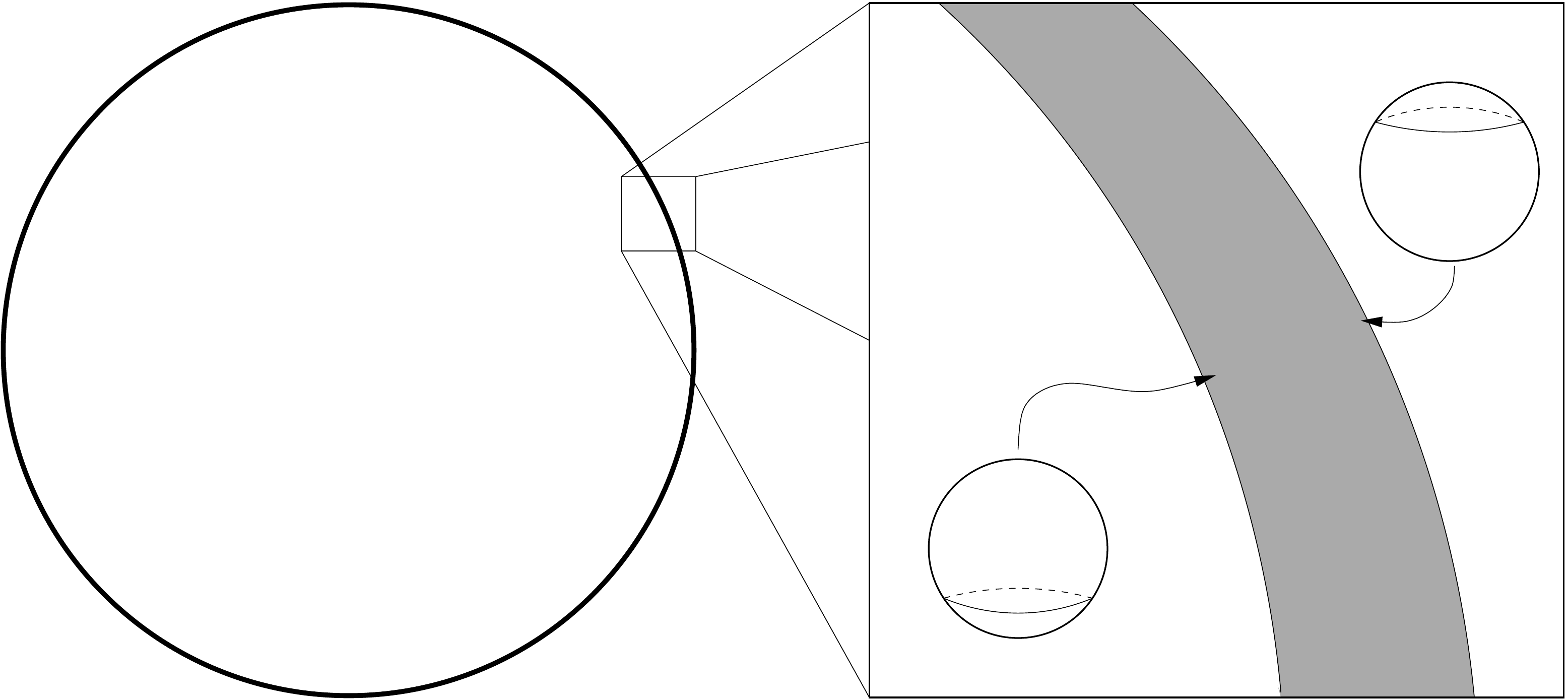} 
\caption{\label{intermediate}The $S^2$ moves across the $S^3$ when you go through the bubble wall. }%
\end{center}
\end{figure}

The combination of the nucleation of the spherical bubble inside spacetime and the Myers effect inside the compact dimensions is visualised in picture \ref{intermediate}. 
Consider some observer in noncompact spacetime that is about to be hit by an expanding BT bubble wall. In reality
this wall is not infinitesimally thin, but has a finite thickness. When the front of the wall passes the observer, the spherical 
D$(p+2)$ wrapping a 2-sphere inside the extra dimensions, starts rolling and reaches the South Pole the moment the back 
of the wall passes the observer. The string theory realisation of the finite thickness wall is a pair 
of D$(p+2)$/$\overline{\text{D}(p+2)}$ branes with $M$ D$p$ branes stretched between them. We have that $p$ antibranes end on the outside of the wall while $M-p$ branes end on the inside of the wall. The $M$ branes stretched between the pair are polarised in the internal space as we  discussed.

Whether this process is energetically favourable, perturbatively or nonperturbatively, depends on the model under consideration. At leading order, in a probe approximation, this can be deduced from the energy given by the full D$(p+2)$ action (DBI+WZ). At first sight it seems puzzling that the brane annihilation process can be classical whereas its associated flux decay bubble is nonperturbative. However, as explained in \cite{Danielsson:2014yga} the bubble nucleation can become arbitrarily fast, such that it is effectively a classical process. 

Consider for instance the SUSY AdS$_7$ backgrounds of \cite{Apruzzi:2013yva} built from 
D6 branes, Romans mass and $H_3$ flux. In this case there is no $A$-cycle (since $p=6$) and the $B$-cycle is indeed topologically a 3-sphere. The D6 branes have been shown to polarise into D8 branes but the D8 branes do not move towards the South Pole, instead they are stuck at stable positions inside the internal manifold. This is consistent with the fact that the vacua are SUSY and hence there should not be any decay possible.  However non-SUSY extensions of this model are possible, both compact models with AdS vacuum and noncompact Minkowski solutions \cite{Blaback:2011pn, Junghans:2014wda}. In the Minkowski case it has been shown that the D8 polarisation cannot occur \cite{Bena:2012tx} and instead these vacua are expected to decay via the NSNS channel which we turn to in the next section.

\section{The decay of NSNS flux} \label{NSdecay}
To understand the decay of NSNS 3-form flux, it is useful to have a simple form of the NS5 brane action. Here we 
solely focus on the Wess-Zumino piece and in the next section we will discuss the details of the DBI action as well. NS5 
worldvolume actions 
have been derived in various places in the literature, see for instance \cite{Eyras:1998hn, Bandos:2000az,Bergshoeff:2011zk} 
(and see \cite{Simon:2011rw} for a useful review on general brane actions and their physics). We will review 
the derivation of the WZ action from first principles given in \cite{Bergshoeff:2011zk} since it gives the WZ action in a form 
slightly  different from the existing literature, which is more useful for our purposes.

\subsection{NS5 Wess-Zumino action}\label{NS5WZactions}
In what follows we use polyform notation in which $C$ is the sum of even RR-form potentials in type IIA and odd in type IIB.\footnote{See \cite{Koerber:2010bx} for a review.}
 We also have worldvolume gauge potentials combined into a polyform $a$. For the IIA/IIB NS5 brane we have
\begin{align}
\text{IIA:}\qquad  &  a =  a_0 + a_2 + a_4 \nonumber \\
\text{IIB:}\qquad  &  a = a_1 + a_3 + a_5 \,,
\end{align}
where the subscript denotes the rank of the form. Our notation is democratic and we naturally incorporate the dual gauge 
potentials as well. Starting from the gauge transformations of the RR potentials
\be
\delta C = \dd \lambda-H_3\w\lambda~,
\ee
where $\lambda$ is a polyform of gauge transformation parameters. We see that the field strengths
\be
F = \dd C-H_3\w C~,
\ee
are gauge invariant. Similarly the gauge transformation of the $B_2$-field, $\delta B_2 = \dd \Lambda_1$ shows that 
the field strength
\be
H_3 = \dd B_2~,
\ee
is gauge invariant. We are interested in the magnetic field strength for $B_2$ under which the NS5 is charged. We 
then take a look at the equation of motion for $B_2$
\be
\dd\left(\e^{-\phi} \star H_3 \right) + \f12 \eta \left\langle F\w \sigma(F) \right\rangle_8 = 0~,
\ee
where $\eta=+1$ in type IIA and $-1$ in type IIB. The angle brackets denote the projection of polyform down to
a specific form degree and $\sigma$ is the reversal operator. We use the definition
\be\label{H7}
H_7 = \e^{-\phi} \star H_3~,
\ee
and demand that it is gauge invariant. By integrating the equation of motion we find an expression for $H_7$ in 
terms of the 6-form gauge potential $B_6$:
\be\label{B6}
H_7 = \dd B_6 - \f12 \eta \left\langle C\w \sigma(F)\right\rangle_7~.
\ee
Gauge invariance then demands
\be
\delta B_6 =\dd \Lambda_5+ \f12 \eta \left\langle \lambda\w\sigma(F)\right\rangle_6~.
\ee
We are now in position to write down a gauge invariant coupling to $B_6$, i.e. we look for a Lagrangian density
${\cal L}_\text{NS5}$ that has the form
\be
{\cal L}_\text{NS5} = B_6 + \cdots
\ee
and is gauge invariant up to a total derivative. To this end we define worldvolume gauge potentials $a$ that transform 
under the RR
gauge transformations
\be
\delta a = \lambda~.
\ee
A gauge invariant field strength can be defined by
\be\label{gaugeinvariant}
{\cal G} = \dd a - H_3\w a - C~,
\ee
where we use the pullbacks of $C$ and $H_3$ onto the worldvolume. With this information one can construct the full gauge invariant action 
\be\label{NS5WZaction}
S_\text{NS5} = \mu_\text{NS5}\int\left\{ B_6 - \f12 \eta \left\langle {\cal G}\w \sigma(C)\right\rangle_6\right\}~.
\ee
The gauge transformation of the Lagrangian density is 
\be
\delta {\cal L}_\text{NS5} = \f12 \eta \dd \left\langle  {\cal G}\w \sigma(\lambda)\right\rangle_5~.
\ee

\subsection{D$p$ annihilation for $p<6$ }

The BT bubble that decays the NSNS flux is an NS5 brane that wraps the $A$-cycle. As discussed above, the BT bubble does not account for the decay of brane charge. We now discuss the microscopic description of the brane charge which proceeds  through an accompanying NS5 brane that wraps a trivial cycle inside the $A$-cycle. The relevant coupling 
in the NS5 brane action is
 \be \label{NScoupling}
-(-1)^p\mu_\text{NS5} \int (\d a_{4-p} - C_{5-p})\w \sigma(C_{p+1})~,
 \ee 
which is there for $p=0,\dots,5$. Note that the factor of half is missing in equation \eqref{NScoupling} compared to eq. \eqref{NS5WZaction}. This is due to the fact that the NS5 worldvolume gauge fields satisfy a self duality relation which halves the eight naive gauge field degrees of freedom living on the brane to the correct four\cite{Bergshoeff:2011zk}. The NS5 brane must wrap a contractible cycle inside the $A$-cycle in order not to have any net NS5 charge (which it cannot have for the sake of charge conservation). This contractible cycle will have dimension $5-p$ such that the NS5 has one codimension inside $A$. Hence the homologically trivial cycle, $\Sigma_{5-p}(x)$, can be labeled by a single real number $x$ denoting the position of the NS5 inside the $A$-cycle.  From (\ref{NScoupling}) we read off that the D$p$ charge induced by the contractible NS5 brane is 
\be
Q(x) = (2\pi)^{\f{p-5}{2}}\int_{\Sigma_{5-p}(x)} (\d a_{4-p} - C_{5-p})\,,
\ee
where the factors of $2\pi$ are obtained by computing $\mu_\text{NS5}/\mu_p$.
We will use a gauge in which
\be
 C_{5-p} (x\to 0) = 0~,
\ee
where $x\to0$ denotes the limit where the contractible cycle shrinks to zero size.
The NS5 induces $\overline{\text{D}p}$ charge $n$ at $x=0$, by having a topological flux of $\d a_{4-p}$ on $\Sigma_{5-p}(x)$:
\be
\int_{\Sigma_{5-p}(x)}\d a_{4-p} = (2\pi)^{\f{5-p}{2}}n \,.
\ee
The way this is achieved is simply by equating $\d a_{4-p} = (2\pi)^{\f{5-p}{2}}n \hat{\epsilon}_{5-p}(x)$, where $\hat{\epsilon}_{5-p}(x)$ is the \emph{normalised} volume-form on $\Sigma_{5-p}(x)$ for any $x$. Since the flux is topological it will pick up a non-zero value even if $\Sigma_{5-p}(x)$ vanishes in size.   The motion from one side of the chain of homological cycles, $x\to0$, to the other side, $x\to1$, corresponds to the decay of $n$ units of $\overline{\text{D}p}$ charge to $M-n$ units of D$p$ charge since
\be
Q (1) =  n - M \,.
\ee
As before, we relied on Stokes theorem
\be
 \int_{x\to1} C_{5-p}-\int_{x\to0} C_{5-p} =  \int_A F_{6-p} = (2\pi)^{\f{5-p}{2}}M~.
\ee

\subsection{D6 annihilation} \label{sec:D6}

An interesting case is the decay of $\overline{\text{D}6}$-branes. From T-duality of the $\overline{\text{D}5}$-brane decay we expect that the decay of NSNS flux in the thick wall limit does not proceed via an NS5 brane that carries $\overline{\text{D6}}$ charge. Also, the NS5 brane does not possess the right couplings to carry $\overline{\text{D6}}$ brane charge. T-duality suggests that the decay uses a KK5 brane carrying $\overline{\text{D6}}$ charge. This poses a problem since a KK5 brane is an object with 6 worldvolume directions, and so, at first sight cannot carry $\overline{\text{D6}}$ charges. However, the KK5 has one special transverse direction that possesses a circle isometry and in essence the spacetime solution looks like a seven-dimensional object. The KK5 worldvolume action also contains the coupling for it to carry $\overline{\text{D6}}$ charge:
\be\label{KKcoupling}
S(\theta) = (n-\theta F_0) \int \iota_k C_7~,
\ee
where $\iota_k$ is the interior product in the direction of the isometry direction of the (Taub-NUT) circle describing the KK5 brane and $\theta$ is the angle along that $S^1$. This coupling can be obtained from massive T-duality of the NS5 coupling term:
\be
\int (n-C_0)\w C_6~.
\ee

\begin{figure}
\begin{center}
\begin{overpic}[width=0.3\textwidth,tics=10]{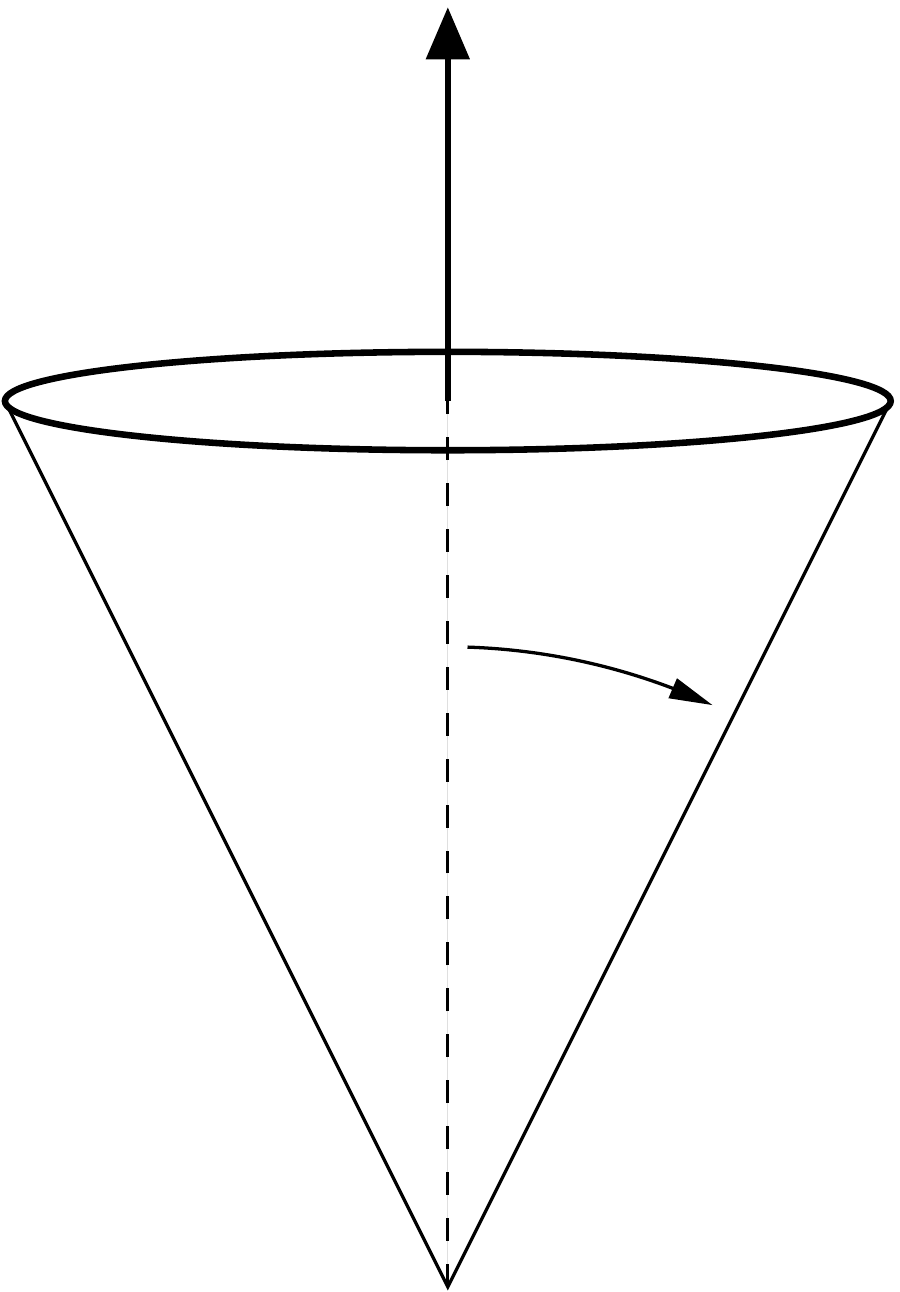}
\put(45,50){$\theta$}
\end{overpic}
\caption{\small\label{cone}The infinitely thick limit of the domain wall KK5-brane that carries D6-brane charge.}
\end{center}
\end{figure}

The question is now what is the spacetime picture of the decay of $H$-flux involving the KK5 brane? The spacetime is seven dimensional Minkowski space which is the worldvolume of the decaying $\overline{\text{D}6}$-branes. The isometry direction of the KK5 must lie in spacetime and so we write the Minkowski metric as
\be
\dd s^2_{\text{Mink}_7} = -\dd t^2 + \dd r^2 + r^2(\dd\theta^2 + \sin^2\theta~\dd \Omega_4^2)~,
\ee
where $\dd \Omega_4^2$ is the metric on the round 4-sphere. The KK5 extends through $(t,r,\Omega_4)$ and $\theta$ parametrises the contractible circle. The codimension of the KK5 in Mink$_7$ is the same as its isometry direction which is nothing but the polar angle in spherical coordinates. As time passes the $\theta$ position of the KK5 evolves from zero to $\pi$ which means that the KK5 has a cone shape (see figure \ref{cone}).
To see that the KK5 carries the correct D6-charge we compare the coupling \eqref{KKcoupling} at both extremum of $\theta$,
\be
S(\pi)-S(0) = -\pi F_0 \int \iota_k C_7 = -F_0\int C_7~,
\ee
where in the last step we used the fact that $C_7$ preserves the symmetry of Mink$_7$ and thus satisfies $\dd \theta\w\iota_k C_7 = C_7$ and is independent of $\theta$.

\begin{figure}
\begin{center}
\begin{overpic}[width=0.4\textwidth,tics=10]{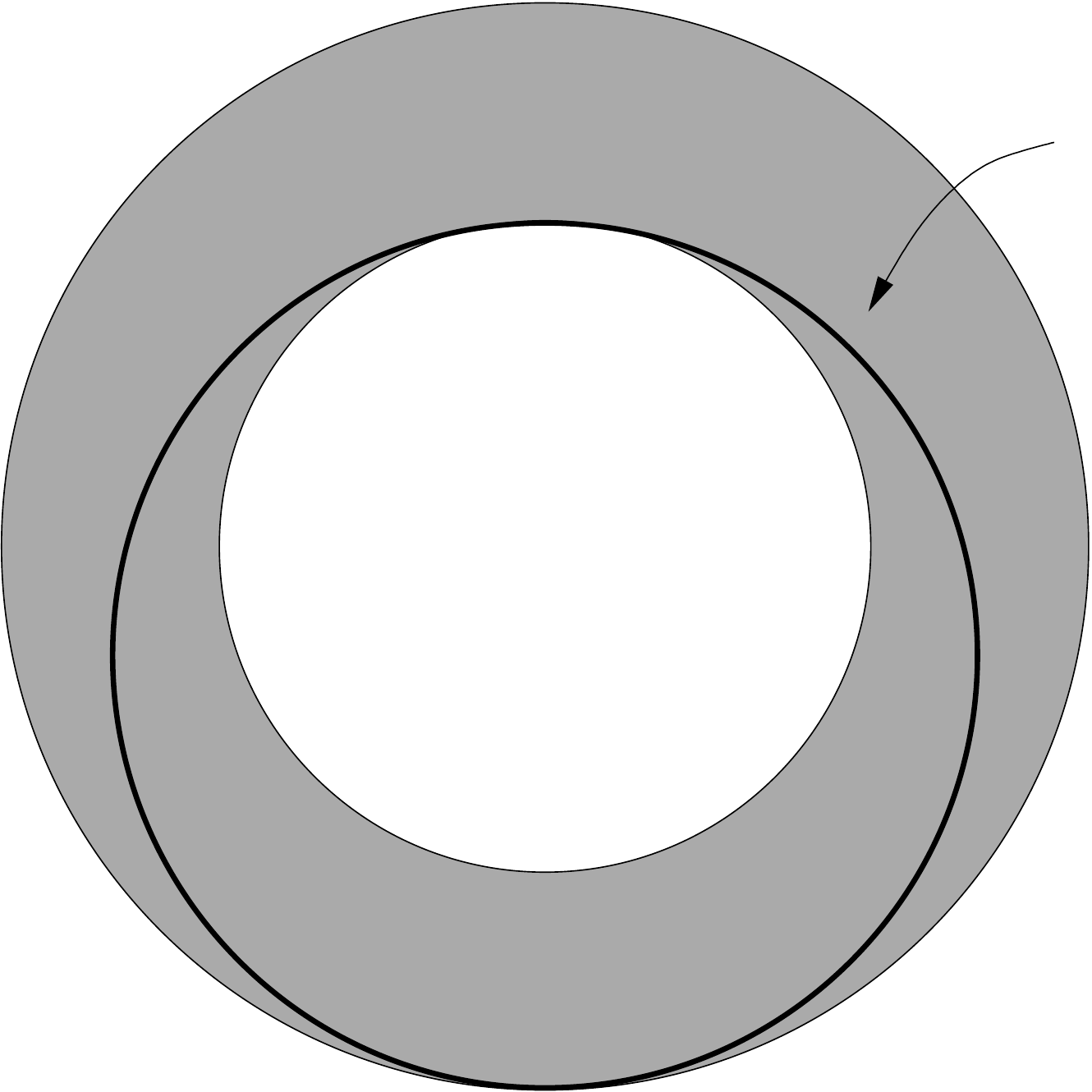}
\put(100,86){KK5}
\end{overpic}
\caption{\small\label{vag}A domain wall NS5-brane with finite thickness that has the associated spherical KK5 in it.}
\end{center}
\end{figure}

The resolved process where a BT bubble carries away $H$-flux \emph{and} D6-charge, must be an NS5 brane of finite thickness with a KK5 flowing inside the wall. We have depicted such a process in figure \ref{vag}.
We do not provide more details here since we plan to present an in-depth study of this decay process in a separate work, which will also address the effect of the backreaction of the $\overline{\text{D6}}$ branes.

\subsection{Tunneling rates}

So far we have not discussed how to compute the actual tunnelling amplitudes for the decay processes. This requires knowledge of the energy it costs to expand a spherical brane or the energy it costs to separate brane/antibrane pairs. This is not captured by the WZ action alone but also involves the DBI action. The DBI actions for the NS5 brane in string frame is
\begin{align}
\text{Type IIA} \quad & -\mu_{\text{NS5}}\int\e^{-2\phi}\sqrt{-\det(g + \e^{2\phi}{\cal G}_1 \otimes{\cal G}_1)}\left(1+\f14 \e^{2\phi}|{\cal G}_3|^2 + \cdots \right)\,,\\
\text{Type IIB} \quad & -\mu_{\text{NS5}}\int\e^{-2\phi}\sqrt{-\det\left(g - \f{\e^{\phi}}{\sqrt{1+\e^{2\phi}{\cal G}_0^2}}{\cal G}_2\right)}\sqrt{1+\e^{2\phi}{\cal G}_0^2}\,,
\end{align}
where the gauge-invariant field strength $\mathcal{G}$ is defined in (\ref{gaugeinvariant}). Note, that these actions are not
written in democratic form as we did for the WZ couplings in section \ref{NS5WZactions} and so one must sometimes dualise 
the field strengths in order to get a DBI action in a suitable form.

Once all interactions are known the potential energy for the spherical NS5 branes, or NS5/$\overline{\text{NS}5}$ pairs, can be computed and hence the tunnelling probability $P$ through the standard WKB or instanton formula \cite{Coleman:1977py,Brown:1987dd}:
\be
P\sim e^{-S_B/\hbar}\,.
\ee
$S_B$ is the instanton action given by
\be
S_B =  \frac{d^d 2\pi^{(d+1)/2}}{(d+1)\Gamma[(d+1)/2] } \frac{T^{d+1}}{\rho^d}
\ee
where $d$ is the number of noncompact dimensions, $T$ is the bubble tension and $\rho$ is the drop in energy density through the bubble (the change in cosmological constant). The brane actions enter in the computation of the tension $T$. For that one constructs first the 1-dimensional effective action that governs the motion of the spherical NS5 brane or the NS5/$\overline{\text{NS}5}$ pair. This action allows classical \emph{domain wall} solutions that interpolate between the two vacua and the on-shell action (tension) of that domain wall equals $T$ (see for instance \cite{Kachru:2002gs}). A classical motion, instead of tunnelling, would correspond to vanishing $T$ and it has been argued that this is possible due to the backreaction of the antibrane probes \cite{Danielsson:2014yga}. 

Instead of computing $T$ using the classical potential derived from the NS5 brane action wrapping some contractible cycle one could simply take the tension of the NS5 domain wall that wraps the \emph{entire} $A$-cycle (the BT one). That tension equals the quantised NS5 tension $g_s^{-2}\mu_{NS5}$ times the volume of the $A$-cycle (see equation \eqref{thinsigma}). The latter way corresponds to the thin wall limit, whereas the first method is the thick wall limit. only in the limit $n/M\ll 1$ can one hope to match the thin and thick limits. But clearly, in the thin wall limit a classical decay can never be spotted.

\section{Illustrations}\label{Illustrations}


\subsection{$\overline{\text{D}p}$-probes in T-duals of the dissolved D6 }

\emph{Fractional brane} solutions in the supergravity context \cite{Herzog:2000rz, Blaback:2012mu} are solutions for D$p$-branes in which there is surrounding flux that carries D$p$ charge as well. This flux is simply the previously mentioned $H_3\wedge F_{6-p}$ combination that contributes to the RR tadpole (\ref{tadpole}). The forces between the flux cloud and the brane sources cancel out when the following condition is obeyed 
\be
\star_{9-p} H_3 = g_s^{(p+1)/4} F_{6-p}\,,
\ee
where the Hodge star is taken with respect to the flat metric on the space transversal to the (parallel) D$p$-branes.  When this condition is met, simple solutions to the equations of motion can be found \cite{Herzog:2000rz, Blaback:2012mu, Kuperstein:2014zda}. For instance for D$6$ solutions in massive IIA SUGRA one has, in Einstein frame \cite{Janssen:1999sa}:
\begin{align}
 \d s^2 &= S^{-1/8}\eta_{\mu\nu}\d x^{\mu}\d x^{\nu} + S^{7/8}[\d r^2 + r^2 \d\Omega_2^2]\,,\nonumber\\
 F_2 &= - g_s^{-3/4}\star_3 \d S\,,\nonumber\\
 F_0 &=  (2\pi)^{-1/2}M \,,\nonumber\\
 H_3 &=  (2\pi)^{-1/2}g_s^{7/4} M r^2 \d r\wedge \d\Omega_2\,,\nonumber\\ 
 e^{\phi} &=g_s S^{-3/4} \,.
\end{align}
The Hodge star is taken with respect to the transversal metric with the warpfactor $S^{7/8}$ taken out.  The function $S$ is given by
\be
S =  1 + \frac{g_sT}{r} -\frac{g_s^{5/2} M^2 r^2}{6}\,,
\ee
where $T$ is the total D6 tension. At sufficiently large $r$ the solution becomes singular, but at small enough distances, which is all we need in this paper, this poses no problem. When there are no D$6$ sources present we can take $T=0$ such that the solution is regular at small $r$. Such (locally) regular solutions can serve as toy models for the local geometries of resolved fractional branes near the tip \cite{Hartnett:2015oda}. 

T-duality of the above solution over the worldvolume directions generates the following solutions
\begin{align}
 \d s^2 &= S^{(p-7)/8}\eta_{\mu\nu}\d x^{\mu}\d x^{\nu} + S^{(p+1)/8}[\d \mathbb{T}_{6-p}^2  +  \d r^2 + r^2 \d\Omega_{2}^2]\,,\nonumber\\
 F_{8-p} &= - g_s^{(p-3)/4}\star_{9-p} \d S\,,\nonumber\\
 F_{6-p} &=  (2\pi)^{(5-p)/2}M \hat{\epsilon}_{6-p}(\mathbb{T}_{6-p})\,,\nonumber\\
 H_3 &= (2\pi)^{(5-p)/2}(-1)^{p} g_s^{(p+1)/4} M r^2 \d r\wedge \d\Omega_2\,,\nonumber\\ 
 e^{\phi} &=g_s S^{(3-p)/4} \,,\qquad S =  1 + \frac{g_sT}{r} -\frac{g_s^{(p-1)/2} M^2 r^2}{6}\,,
\end{align}
where $T$ is the total D$p$ tension and $\mathbb{T}_{6-p}$ is the torus generated by T-duality. We again consider the throat geometries which are locally smooth since $T=0$ and we will focus on the geometry at the tip defined by $S=1$ or $r=0$.

Since it is useful for computing the WZ action for an NS5 brane we compute the dual NSNS gauge potential $B_6$, which is defined in (\ref{B6}). A gauge-fixed solution of this equation, at the tip ($S=1$), is:\footnote{In the potential calculations below we will make use of a slightly different expression, which only makes
sense for the pullback of $B_6$ on to the NS5 worldvolume. In stead of using \eqref{BB6} we will use
\[
B_6 = g_s^{(p-3)/4}\d x^0\wedge \ldots\wedge \d x^p \wedge {\cal G}_{5-p}~,
\]
where ${\cal G}_{5-p}$ is defined in \eqref{gaugeinvariant}. This shifts $B_6$ by a constant, and hence is just another gauge choice for it.}
\be \label{BB6}
B_6 = -g_s^{(p-3)/4}\d x^0\wedge \ldots\wedge \d x^p \wedge C_{5-p}~.
\ee
In this way the $g_s$ dependence of the WZ action matches that of the DBI part in Einstein frame, as expected. In the same gauge we have that, at the tip,
\be
C_{p+1}(r=0)=0\,.
\ee

In what follows we consider perturbations of these local geometries by $n$ $\overline{\text{D}p}$ probes. The following conditions are necessary for having a SUGRA description and a probe limit at the same time: 
\be
n\ll M\,,\qquad g_sn\gg 1\,,\qquad g_s\ll 1 \,.
\ee
The first condition makes sure that the probe approximation is naively justified, the second condition ensures that the \emph{Schwarzschild radius} of the probe is large in string units in order to suppress higher derivative corrections and the last condition is the validity of perturbative string theory.

Below we verify how the probe branes can annihilate with one unit of $H$-flux to become $M-n$ D$p$ branes. With regard to the previous sections the $A$-cycle in these backgrounds are the T-duality circles $\mathbb{T}^{6-p}$ and the $B$-cycle is noncompact.

\subsection{Brane nucleation on tori}

In the literature on antibranes polarising into larger objects over compact cycles (for instance \cite{Kachru:2002gs, Klebanov:2010qs}) the compact cycles tend to have sphere topologies. The prime example comes from $\overline{\text{D3}}$ branes inside the KS throat that polarise into NS5 branes wrapping a contractible $S^2$ inside an $S^3$ (the $A$-cycle). The $S^2$ pinches off at the poles and the motion between the two poles corresponds to the annihilation of branes and fluxes \cite{Kachru:2002gs}. But when the compact cycles are toroidal, or have any other topology different from a sphere some extra complications arise. 

\begin{figure}
\begin{center}
\begin{overpic}[width=0.3\textwidth,tics=10]{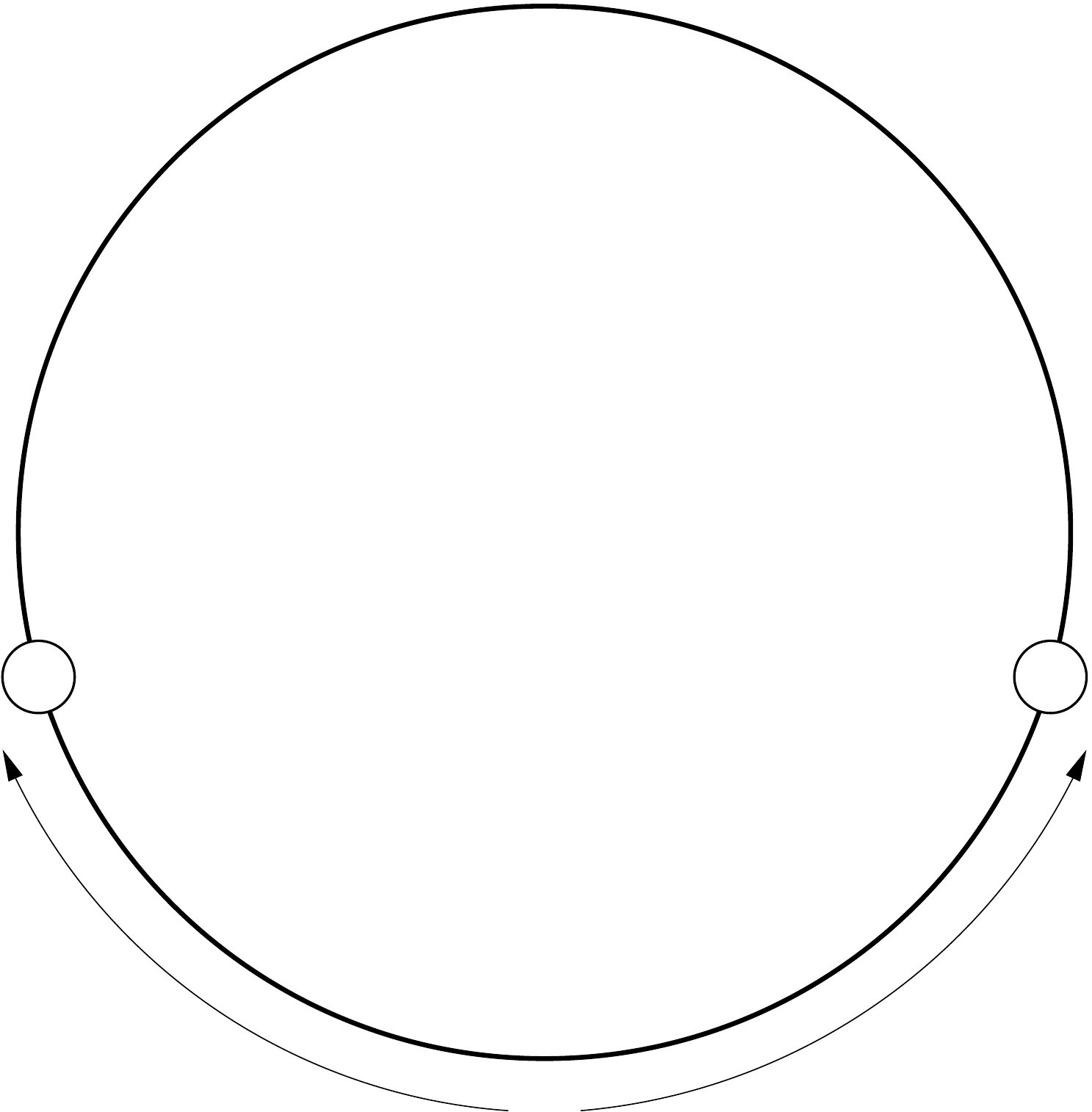}
\put(103,37){NS5}
\put(-20,37){$\overline{\text{NS5}}$}
\end{overpic}
\caption{\label{1torus} Nucleation of a NS5/$\overline{\text{NS5}}$ pair carrying D5 charges. }%
\end{center}
\end{figure}

Let us start with a compact $A$-cycle that is $S^1=\mathbb{T}^1$, as is the case for $p=5$. Then the NS5 brane carrying $\overline{\text{D5}}$ charge necessarily is an NS5/$\overline{\text{NS5}}$ pair as in picture \ref{1torus}. 
We take the circle of length 1 such that the circle-coordinate $\theta$ ranges from  $(-1/2, 1/2)$. The profile for $C_0$ is then
\be
C_0 = M\theta\,.
\ee
Consider the nucleation of an NS5/$\overline{\text{NS5}}$ pair at $\theta=0$. The pair separates and moves to opposite sides and meet again at $\theta=1/2$. The NS5's carry worldvolume flux $n/2$ and $-n/2$. The total induced charge is 
\be
Q(\theta) \sim \Bigl(\frac{n}{2} - M\theta  \Bigr) - \Bigl(-\frac{n}{2} + M\theta  \Bigr) = \Bigl( n - 2M\theta  \Bigr)\,.
\ee
The two WZ actions are subtracted because one WZ term corresponds to an $\overline{\text{NS5}}$. The second WZ action also has a different sign for the $M\theta$ term because it moves in the opposite direction. So $\theta$ here is the positive $\theta$ that tracks the NS5 and not the $\overline{\text{NS5}}$ that is at the opposite side. So the motion ranges from $\theta=0$ to $\theta=1/2$ such that we indeed find that $Q(\theta)$ ranges from $n$ to $(n-M)$. 

\begin{figure}
\begin{center}
\includegraphics[height=40mm]{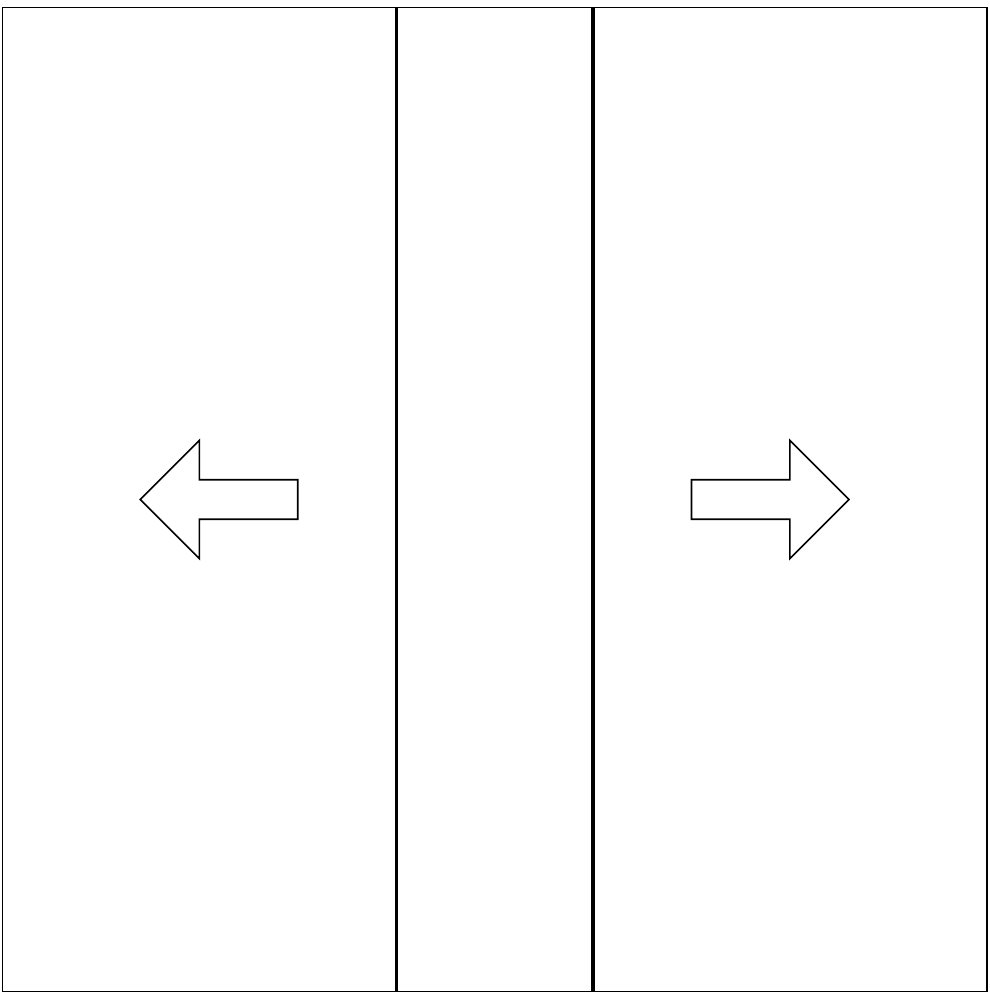} \qquad \includegraphics[height=40mm]{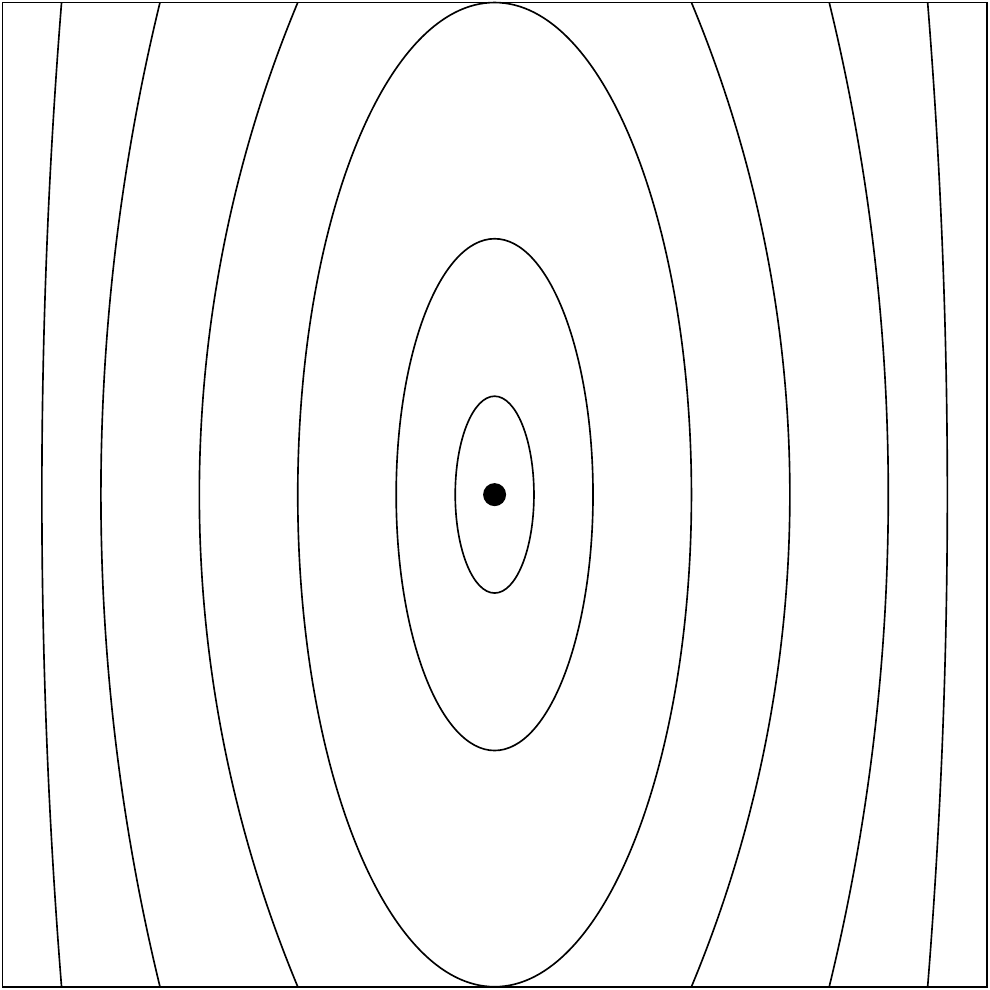} \qquad \includegraphics[height=40mm]{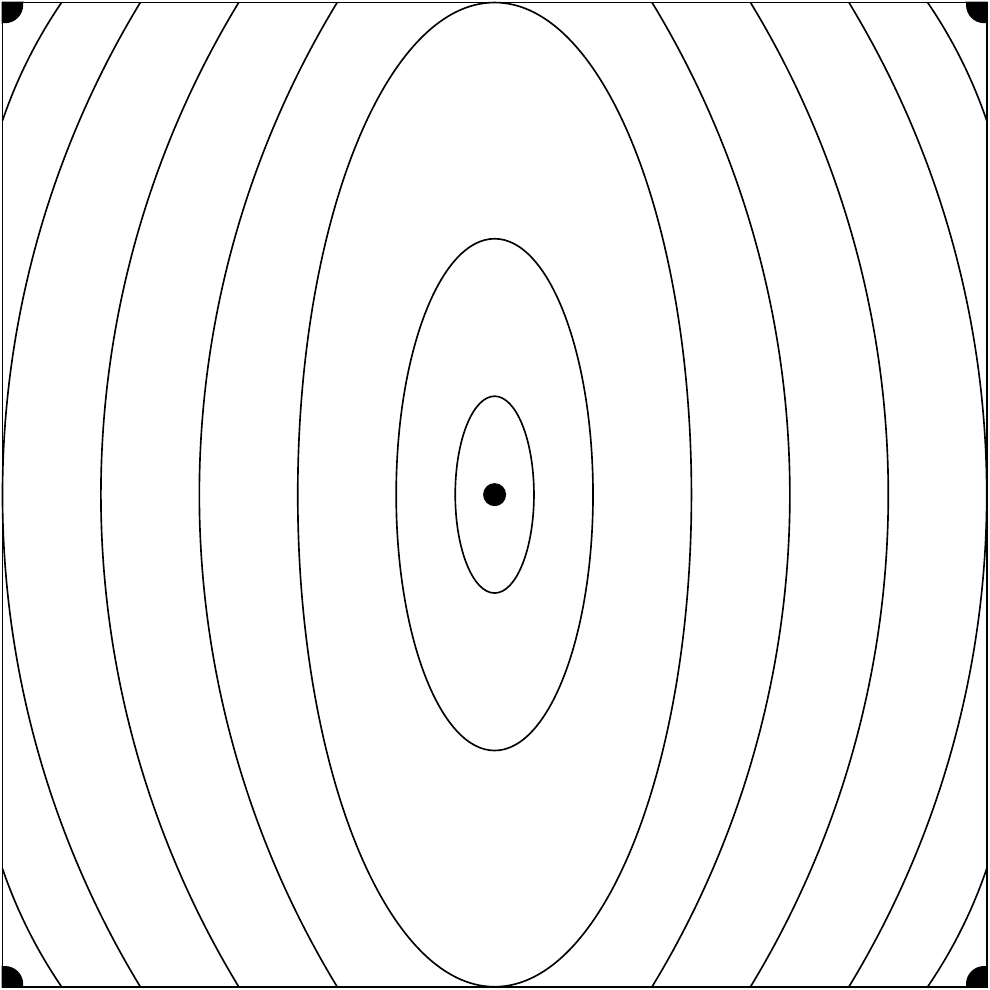} 
\caption{\small \label{torus}  NS5 nucleation on tori. The tori are respresented as squares with the opposite sides identified. Left picture: Nucleation of an NS5/$\overline{\text{NS5}}$ pair  that moves in opposite signs until they meet again.  Middle picture: Nucleation of a spherical NS5 that transitions into an NS5/$\overline{\text{NS5}}$ pair or vice versa. Right picture: nucleation of a growing spherical NS5 that shrinks again as a spherical NS5 on a different part of the torus, by making a transition into an NS5/$\overline{\text{NS5}}$ pair.}%
\end{center}
\end{figure}

Direct T-duality suggest that the other branes decay via NS5/$\overline{\text{NS5}}$ pairs wrapping $5-p$ circles inside $\mathbb{T}^{6-p}$. This is correct but not unique. Similar to cycles with a spherical topology, we can have spherical NS5 branes nucleating inside the torus. Spherical NS5 branes are morally NS5/$\overline{\text{NS5}}$ pairs. We expect two possible things to occur for spherical NS5 branes carrying $n$ $\overline{\text{D}p}$ charges: 
\begin{enumerate}
\item The spherical NS5 grows and intersects itself in some, but not all directions of the torus to create a NS5/$\overline{\text{NS5}}$ pair that eventually decays by collision.
\item The spherical NS5 grows and intersects itself to form an NS5/$\overline{\text{NS5}}$ pair that  eventually reconnects again into a single spherical NS5 brane that shrinks to a single point at a different position in the torus than it started off. 
\end{enumerate}
We have depicted all the possible processes for a 2D torus in figure \ref{torus} .

The process in which a spherical NS5 breaks into an NS5/$\overline{\text{NS5}}$ pair wrapping non-contractible cycles, or the reversed process, is expected to happen through tachyon condensation. Some details of this can be found in \cite{Kleban:2011cs,DAmico:2014psa}.

\subsection{$\overline{D3}$ probes}

For the sake of analogy with the work of Kachru, Pearson and Verlinde \cite{Kachru:2002gs} we consider the situation in which a spherical NS5 nucleates and collapses as a spherical NS5 in a different part of the torus. As explained above, this occurs via breaking the NS5 sphere into NS5/$\overline{\text{NS5}}$ pairs that later combine again into a single NS5 sphere.

Consider Cartesian coordinates on the straight torus with unit lengths.  If we take the $\overline{\text{D3}}$ to polarise at the position $S=(0,0,0)$ into a round spherical NS5 then it will collapse again at position $N=(1/2, 1/2, 1/2)$.  When the $\mathbb{T}^3$ is replaced by $\mathbb{R}^3$ the computation of the NS5 potential was performed in \cite{Hartnett:2015oda} and for a spherical NS5-brane that starts in the center of the $\mathbb{R}^3$ the polarisation potential reads
\be
V\sim\sqrt{ \f{r^4}{(2\pi)^2} + \Bigl( n - \frac{M}{3}r^3\Bigr)^2} +\left(n- \frac{M}{3}r^3\right)\,,
\ee
where $r$ is the radial coordinate in the local patch with $ S=(0,0,0)$ as the center of the radial coordinates.  This potential has a local minimum at finite $r$ and for $n$ small enough we can make sure this minium occurs in a region of validity of the coordinate system.  For large $r$ it simply keeps on rising so for a noncompact $A$-cycle given by $\mathbb{R}^3$ decay via NS5 nucleation does not occur.  To understand whether the potential goes down at some point on the torus it suffices to compute the potential at the point where one expects spherical collapse and verify whether the potential is indeed lower then at the starting point $S=(0,0,0)$. If this is the case, one expects the nonperturbative transition to occur, even though the full potential is not known. The point of spherical collapse in this example is engineered to be $N=(1/2, 1/2, 1/2)$, but the transition to the spherical collapse cannot be described in the same spherical coordinate system.  We therefore consider the radial coordinates around the point $N$ and denote the radial coordinate in that patch as $\tilde{r}$. To compute the potential at $N$ we require the value of $C_2$ at that point which then fixes $B_6$ through (\ref{BB6}).
We already argued that in the gauge where
$\int_{S} C_2=0$, then $\int_N C_2 =2\pi M$. Straightforward calculation leads to the potential in the $\tilde r$-coordinates,
\be
V\sim\sqrt{ \f{\tilde r^4}{(2\pi)^2} + \Bigl( n  - M+ \frac{M}{3}\tilde r^3\Bigr)^2} +\left(n-M+ \frac{M}{3}\tilde r^3\right)\,.
\ee
We find that indeed the potential energy at $N$ is lower than at $S$:
\be
V(\tilde{r}=0) = 0 \quad<\quad V(r=0) = 2 n 
\ee

Hence we notice something very different from the noncompact $A$-cycle: the minimum of the brane polarisation at small $r$ is only a local minimum since the potential at $N$ is smaller then the potential at the minimum. At $N$ the background is BPS again and the antibranes annihilated against the 3-form flux, in complete parallel with $\overline{\text{D3}}$ branes at the tip of the KS throat \cite{Kachru:2002gs}.

\subsection{$\overline{\text{D4}}$ probes}

Now we consider the T-dual process on a 2-torus.  The blow up of D4 branes into NS5 branes has been considered earlier \cite{Bena:2002wg}, but not in the background described here. As before there are multiple options. We need to compute $C_1$, which also fixes $B_6$ via (\ref{BB6}), to understand this process. To describe the growth of a spherical NS5 we consider local radial coordinates on the torus such that, around $N$ 
\be
C_1 = \frac{\sqrt{2\pi}M}{2}r^2\d\phi\,.
\ee
This is sufficient information to plug into the DBI and WZ action to get an expression very similar to what we found for the $\overline{\text{D3}}$ above. We hen ce find a local minimum for the spherical NS5 radius inside $\mathbb{T}^2$. Since the computation and expression for the spherical NS5 is in complete analogy with the $\overline{\text{D3}}$ example we do not bother the reader with the explicit expressions.

Alternatively one can consider an NS5/$\overline{\text{NS5}}$ pair wrapping an $S^1$ inside $\mathbb{T}^2$, move in opposite directions and annihilate to leave $M-n$ D4 branes at the other side. To find the potential for that process one uses Cartesian coordinates $(x,y)$ and the following expression for $C_1$
\be
C_1 =\sqrt{2\pi}M x\d y\,,
\ee
The motion of the NS5 branes is from $x=0$ to $x=1/2$ where it meets the $\overline{\text{NS5}}$ that came from the other side. The potential for the process would therefore be a function of $x$. The way to compute the potential is identical to what we describe below for $\overline{\text{D5}}$ branes.

\subsection{$\overline{\text{D5}}$ probes}
Anti-D5 probes can only decay in the NSNS channel by transforming into an NS5/$\overline{\text{NS5}}$ pair since there is no option for spherical subcycles inside $S^1$. We already described above how the charge evolves from $n$ to $n-M$ and in what follows we compute the difference in energies between the two configurations.

\begin{figure}
\begin{center}
\includegraphics[width=0.6\textwidth]{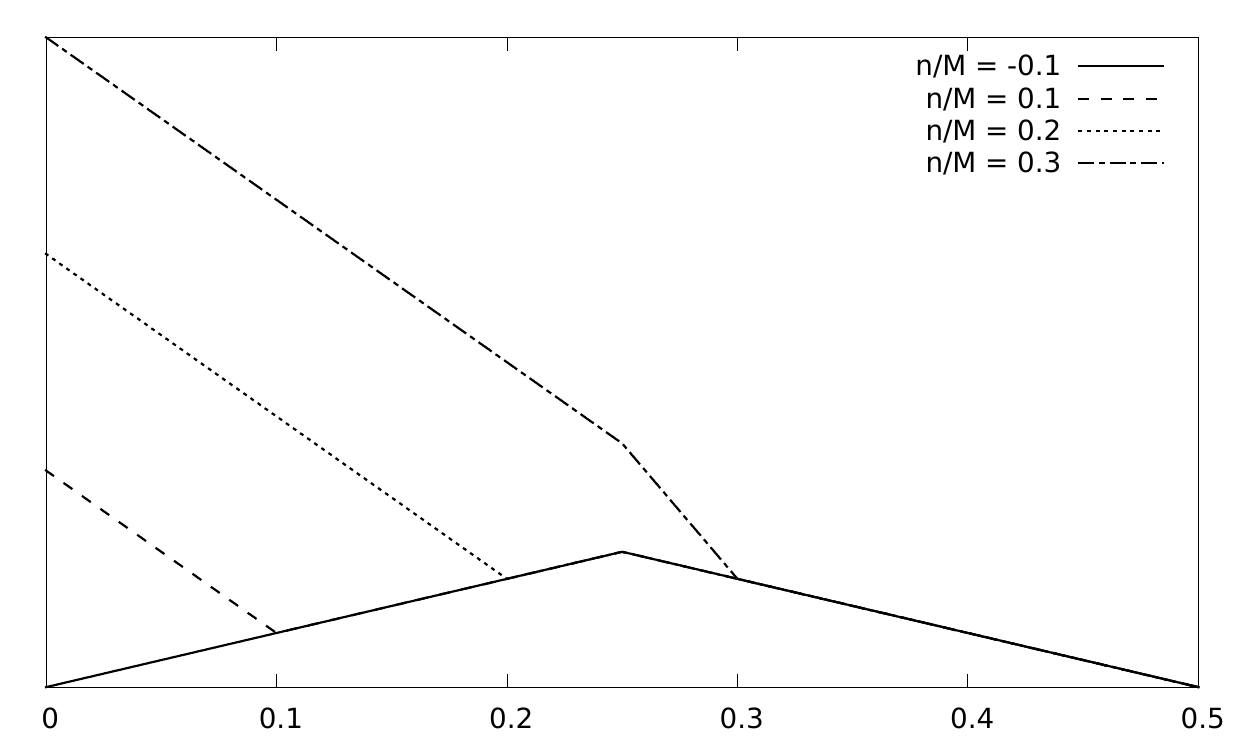} 
\caption{\small \label{D5potential}  The brane-flux decay potential for $\overline{\text{D5}}$ brane plotted for various values of $n/M$. As expected for small enough $n/M$ a metastable state appears.}%
\end{center}
\end{figure}

As before we take $C_0 = M\theta$ and the NS5/$\overline{\text{NS5}}$ pair at $\theta=0$. The NS5's moves from $0$ to opposite sides and meet at $\theta=1/2$. The NS5's carry worldvolume flux $n/2$ and $-n/2$. The total energy can be found from summing the worldvolume actions and  taking into account the mutual interaction term. This is technically different from spherical NS5 branes in which the energy that it costs to expand the NS5 is taken into account already by the DBI term. Instead the energy it costs to overcome the electric attraction between an NS5/$\overline{\text{NS5}}$ pair must be added, just like in a Schwinger pair-creation computation. A reasonable guess is that this potential is linear in the separation between the brane and antibrane as follows
\be
V_\text{interaction}\sim -|\theta -\tfrac{1}{4}|
\ee
where we normalise the potential such that it has an unstable fixed point at $\theta=1/4$.\footnote{This linear behavior agrees with dimensional counting, but the worried reader might want to know that the form of this term does not affect the energy difference between the starting point and the end point since at both these points the binding energy of the NS5/$\overline{\text{NS5}}$ pair should be identical.} The (normalised) energy at either $\theta=0$ ($N$) or $\theta=1/2$ ($S$) reads
\be
V = 2g_s^{-1/2}\sqrt{1+g_s^2\mathcal{G}_0^2} - 2g_s^{1/2}{\cal G}_0 \approx Mg_s^{1/2}\left(\left|\f n{M} - 2\theta\right|+\Big(\f nM-2\theta\Big)\right)\,,
\ee
where we threw away terms of order $(g_sM)^{-1}$. If we include the interaction term $-Mg_s^{1/2}(|1/4-\theta|-1/4)$ we can plot the full potential
\be
V = Mg_s^{1/2}\left(\left|\f n{M} - 2\theta\right|+\left(\f n{M} - 2\theta\right) -\left|\f14-\theta\right|+\f14\right)
\ee
 over the entire circle and the result is shown in figure \ref{D5potential} for various values of $n/M$. As expected for small enough $n/M$ a metastable state appears. At the endpoints of the circle the potential takes the value (for $p/M>0$)
 \be
V(\theta=0) = 2g_s^{1/2}n~\,,\qquad V(\theta=1/2) = 0~.
 \ee

We see that the potential at $\theta=0$ gives twice the tension and zero at the other point which is expected since the fluxes and branes are mutually BPS. 
\newpage

\section{Discussion}\label{Discussion}

We have explained how brane-flux annihilation is expected to proceed in flux compactifications. A useful lead is the RR tadpole condition (\ref{tadpolecondition}) which implies that flux decay has to come together with brane decay. Perhaps confusingly, this implies that there are two contractible branes that mediate the decay: the Brown-Teitelboim bubble in spacetime and the polarised brane wrapping a contractible cycle in the internal manifold. We argue that this should be viewed as two distinct limits of the same process. In this way the polarised brane can be seen as the stringy resolution of the thin BT bubble as was also argued in \cite{Danielsson:2014yga}. 

Apart from explaining the general physical picture we have found a new family of local throat geometries in which antibranes can form metastable states in the probe approximation. These throat geometries have toroidal cycles, and to our knowledge, it was not yet explained how brane-flux decay proceeds in this case. 

Since the discovery \cite{Bena:2009xk, McGuirk:2009xx, Gautason:2013zw} that the backreaction of antibranes leads to singular $H_3$ and $F_{6-p}$ fluxes (in addition to the normal $F_{p+2}$ singularity) it has been questioned whether the backreaction invalidates the probe approximation and can  cause perturbative instabilities  \cite{Bena:2012ek, Blaback:2012nf, Bena:2014jaa, Blaback:2014tfa, Danielsson:2014yga,Danielsson:2015eqa, Bena:2015kia}, but see also \cite{Michel:2014lva} \cite{Hartnett:2015oda} for arguments against this interpretation.  What obstructs a clear argument to settle the debate is a solution of a localised antibrane and an understanding of its decay channel in the SUGRA limit. For anti-D6 branes however, the localised solutions are known \cite{Blaback:2011pn} but the decay channel was not. In this paper we speculate that the thick wall NSNS decay channel proceeds via KK5 nucleation. If correct, it should be possible to verify whether the backreaction indeed causes the background to decay perturbatively. This can be done consistently by considering a probe anti-D6 brane in the background of $n-1$ backreacting anti-D6, such that still $n\ll M$ required for metastability in the probe picture. We hope to report on this in the near future. 

We can try to use the results of this paper to explain the findings in \cite{Hartnett:2015oda}, where it was claimed, contrary to \cite{Bena:2012ek, Blaback:2014tfa}  that smooth finite temperature solutions can exist, which is an argument in favour of metastable states beyond the probe approximation. The crucial issue is that the $A$-cycle used in the smooth finite temperature solutions of \cite{Hartnett:2015oda} is noncompact. 
By increasing the size of the torus in the toroidal antibrane examples of section \ref{Illustrations}, one can make the classical barrier against brane-flux decay arbitrarily large (in the probe approximation). Hence in the limit $\mathbb{T}^{6-p}\rightarrow \mathbb{R}^{6-p}$ brane-flux decay becomes impossible. This is consistent with the findings of \cite{Hartnett:2015oda} that for noncompact $A$-cycles of the form $\mathbb{R}^{6-p}$ smooth finite temperature solutions exist. Since in that limit there is no brane-flux decay and the viewpoint of \cite{Blaback:2012nf, Danielsson:2014yga} that explains the absence of smooth finite temperature solutions does simply not apply (and in absence of a noncompact $A$-cycle neither does the nogo-theorem \cite{Blaback:2014tfa} for existence of smooth solutions apply).

\subsection*{Acknowledgements}
We  thank Iosif Bena and Ulf Danielsson for useful discussions and Marjorie Schillo for many helpful comments on the manuscript.
TVR is supported by the Pegasus and Odysseus programme of the FWO.  BT is aspirant FWO. We acknowledge support from the European Science Foundation Holograv Network.

\newpage
\appendix

\section{Different types of brane-flux decay}\label{Generalised}
The discussion so far has focused on preserving the RR tadpole through a domain wall bubble nucleation, and we have argued that this is accomplished by internal dynamics involving D-branes, NS5-branes or a KK5-brane. This is relevant for compactifications/solutions where only one RR flux is turned on together with $H$-flux. In two dimensions we may encounter multiple RR-fluxes for which the NSNS tadpole becomes non-trivial as can be seen from the integrated $B_2$-equation of motion:
\be
\f12 \eta \int\left\langle F\w \sigma(F)\right\rangle_8 = 2\kappa_{10}Q_{\text{F}1}~,
\ee
where on the right hand side the total fundamental string charge enters. The F1 sources wrap the two dimensional spacetime but cannot appear in 
higher dimensions without breaking Lorenz invariance of spacetime.  The decay of flux numbers in this case must involve the polarization 
of F$1$'s to a relevant D-brane that flows across compact cycles in the internal space to give rise to decay of F$1$ charge (see for instance \cite{Kasai:2015exa}).

Although it is well known that D-branes source  the fundamental string, it is useful to observe this effect from the worldvolume point of view. 
In our conventions the full WZ terms of the D$p$-brane 
worldvolume action takes the form
\be
(-1)^p\mu_{p}\int \langle \sigma(\e^{\cal F}\w C)\rangle_{p+1} 
\ee
where ${\cal F}_2=F_2 - B_2$ as before. Previously, we discussed how D$p$-branes can polarize to a D$(p+2)$ brane via the standard Myers effect. The worldvolume gauge field of the D$(p+2)$ brane takes a monopole configuration that generates 
the brane charge of the original brane. But when a D-brane sources F$1$ charge, the worldvolume field strength ${\cal F}_2$ also takes a nontrivial
configuration which is \emph{electric} rather than magnetic. ${\cal F}_2$ is then such that the that the combination
\[
\langle\e^{\cal F}\w C\rangle_{p-1}~,
\]
threads a topologically trivial 
$p-1$ cycle. The topology of the D$p$ brane is that of the F$1$ it is sourcing times a topologically trivial $p-1$ cycle. Using the standard
DBI action and the WZ terms written above one can derive the potential for such a brane in a given background, this was the approach taken in 
\cite{Klebanov:2010qs}. Coupling the WZ action above with the bulk kinetic term for the $B$-field we find the equation of motion
\be
\dd H_7 + \f12 \eta \left\langle F\w \sigma(F)\right\rangle_8 = 2\kappa_{10}(-1)^p\mu_p\langle\sigma(\e^{\cal F}\w C)\rangle_{p-1}\w\delta_{9-p}~.
\ee
The F1 charge of the D-brane is measured by the right hand side of this equation. Again we can motivate that the F1 charge
changes as the D-brane travels across a compact $p-1$ cycle. Taking only the leading order term of the above coupling and assuming a spherical
$p-1$ cycle for simplicity we see that
\be
Q(\Psi=\pi)-Q(\Psi=0) \propto \int_{\Psi=\pi} C_{p-1}-\int_{\Psi=0} C_{p-1}=  \int F_p~.
\ee
The accompanying BT bubble is a Domain wall D-brane of the same dimension as the brane that sources the F1 charge. So
in summary the annihilation of RR-fluxes with fundamental string is closely related to the annihilation of RR-fluxes with D-branes discussed in
section \ref{sec:RRdecay}, the only difference is that the polarized brane that travels in the internal space has an electric worldvolume configuration in stead of 
the magnetic one discussed in section \ref{sec:RRdecay}. Otherwise the physics of the annihilation is completely analogous.

We have not covered all possible decay processes. But we emphasized those that are relevant for the most typical flux compactifications. More general possibilities can for instance involve supertubes \cite{Bena:2011fc}.

\bibliographystyle{utphys}
{\footnotesize
\bibliography{refs}} 
\end{document}